\documentclass[%
notitlepage,nofootinbib,superscriptaddress,
showpacs,preprintnumbers,
amsmath,amssymb,
prd,onecolumn%
]{revtex4-1}

\usepackage{epsfig}
\usepackage{graphicx}
\usepackage{bbm}
\usepackage[colorlinks=true,citecolor=blue]{hyperref}
\draft 
\begin{document}
\title{Near Maximal Atmospheric Neutrino Mixing in Neutrino Mass Models with Two Texture Zeros}
\author{S. Dev}
 \email{dev5703@yahoo.com}
    \affiliation{Department of Physics, Himachal Pradesh University, Shimla 171005, INDIA.}%
    \affiliation{Department of Physics, School of Sciences, HNBG Central University, Srinagar, Uttarakhand 246174, INDIA.}
\author{Radha Raman Gautam}%
 \email{gautamrrg@gmail.com}
\affiliation{ 
Department of Physics, Panjab University, Chandigarh 160014, INDIA.}%
 \author{Lal Singh}%
 \email{lalsingh96@yahoo.com}
  \affiliation{%
   Department of Physics, Himachal Pradesh University, Shimla 171005, INDIA.}%
\author{Manmohan Gupta}%
 \email{mmgupta@pu.ac.in}
\affiliation{ 
Department of Physics, Panjab University, Chandigarh 160014, INDIA.}
\title{Near Maximal Atmospheric Neutrino Mixing in Neutrino Mass Models with Two Texture Zeros}
\begin{abstract} The implications of a large value of the effective Majorana neutrino mass for a class of two texture zero neutrino mass matrices have been studied in the flavor basis. It is found that these textures predict near maximal atmospheric neutrino mixing angle in the limit of large effective Majorana neutrino mass. It is noted that this prediction is independent of the values of solar and reactor neutrino mixing angles. We present the symmetry realization of these textures using the discrete cyclic group $Z_3$. It is found that the texture zeros realised in this work remain stable under renormalization group running of the neutrino mass matrix from the seesaw scale to the electroweak scale, at one loop level.
\end{abstract}
\pacs{14.60.Pq, 11.30.Hv, 14.60.St}
\maketitle
\section{Introduction}
The flavor mixing pattern in the lepton sector is quite different from the mixing pattern in the quark sector. In the quark sector, one of the mixing angles is around $13^\circ$ and the other two mixing angles are very small whereas in the lepton sector two of the mixing angles (atmospheric mixing angle $\theta_{23}$ and solar mixing angle $\theta_{12}$) are large and the third mixing angle (reactor mixing angle $\theta_{13}$) which was recently measured in a number of neutrino oscillation experiments~\cite{t2k,minos,dchooz,dayabay,reno} is around $9^\circ$. To explain the mixing pattern in the lepton sector, a number of theoretical ideas have been proposed. A particular approach for explaining the lepton flavor mixing pattern is based on non-Abelian discrete symmetries which predict values of mixing matrix elements independent of the lepton masses and are known as mass independent textures. Some typical mixing patterns obtained using the above approach are tribimaximal mixing~\cite{tbm}, bimaximal mixing~\cite{bm}, golden ratio-I~\cite{gr1}, golden ratio-II~\cite{gr2}, hexagonal mixing~\cite{hm} all of which predict a vanishing reactor mixing angle. However, in the light of recent experimental data~\cite{t2k,minos,dchooz,dayabay,reno}, these mixing patterns need modifications. The other approaches used to explain lepton flavor mixing relate mixing matrix elements to lepton masses include texture zeros~\cite{fgm,tz,ourdegeneracies,xingtz,peinado}, vanishing minors~\cite{zmlashin, zerominor,ourzm}, hybrid textures~\cite{hybrid}, equalities between the elements~\cite{tec}.\\
Two texture zeros in the effective neutrino mass matrix ($M_\nu$) in a basis where the charged lepton mass matrix ($M_l$) is diagonal have been extensively studied in the past~\cite{fgm,tz,ourdegeneracies,xingtz,peinado}. Out of the 15 possible cases of two texture zeros in $M_\nu$, only 7 are compatible with the present neutrino oscillation data. The seven allowed cases of two texture zeros in the nomenclature of Ref.~\cite{fgm} are listed in Table~\ref{tab1}.\\
\begin{small} 
\begin{table}[htp]
\begin{center}
\begin{tabular}{|c|c|c|c|}
\hline $A_1$ & $A_2$ & $B_1$ & $B_2$ \\ 
 \hline $\left(
\begin{array}{ccc}
0 & 0 & \times \\ 0 &\times & \times \\ \times & \times & \times
\end{array}
\right)$ & $\left(
\begin{array}{ccc}
0 & \times & 0 \\ \times & \times & \times \\ 0 & \times & \times
\end{array}
\right)$ & $\left(
\begin{array}{ccc}
\times & \times & 0 \\ \times & 0 & \times \\ 0 & \times & \times
\end{array}
\right)$ & $\left(
\begin{array}{ccc}
\times & 0 & \times \\ 0 & \times & \times \\ \times & \times & 0
\end{array}
\right)$  \\ 
\hline $B_3$ & $B_4$ & $C$ & - \\ 
\hline $\left(
\begin{array}{ccc}
\times & 0 & \times \\ 0 &0 & \times \\ \times & \times & \times
\end{array}
\right)$ & $\left(
\begin{array}{ccc}
\times & \times & 0 \\ \times & \times & \times \\ 0 & \times & 0
\end{array}
\right)$ & $\left(
\begin{array}{ccc}
\times & \times & \times \\ \times & 0 & \times \\ \times & \times & 0
\end{array}
\right)$ & -   \\ \hline
\end{tabular}
\caption{\label{tab1}Viable two texture zero neutrino mass matrices. $\times$ denotes the non-zero elements.}
\end{center}
\end{table}
\end{small}
It was shown by Grimus et al.~\cite{grimus} that classes $B_3$ and $B_4$ of two texture zeros predict a near maximal atmospheric neutrino mixing angle when supplemented with the assumption of quasidegenerate neutrino masses and this prediction is independent of the values of solar and reactor neutrino mixing angles. It was found by Dev et al.~\cite{ourzm} that a near maximal $\theta_{23}$ is predicted for two classes ($B_5$ and $B_6$ in the classification scheme of Ref.~\cite{zmlashin}) of two vanishing minors in $M_\nu$ in the limit of a large value of the effective Majorana neutrino mass ($|M_{ee}|$).\\ In the present work, a near maximal $\theta_{23}$ has been predicted for classes $B_1$, $B_2$, $B_3$ and $B_4$ of two texture zeros in the limit of a large $|M_{ee}|$ and this prediction is independent of the values of solar and reactor neutrino mixing angles. The symmetry realization of these texture structures using $Z_3$ symmetry has been presented. It has been shown that the texture zeros realised in this work remain stable under renormalization group (RG) running of the effective neutrino mass matrix at one loop level.\\
The paper has been organised as follows: Section 2 describes the framework used to obtain the constraint equations for two texture zeros. In Section 3, we give the details of numerical analysis. Section 4 is devoted to the symmetry realization of the two texture zero classes considered in this work. In section 5 we discuss the stability of two texture zeros under RG running. Section 6 summarizes this research work.
\section{The Framework}
In the flavor basis the complex symmetric neutrino mass matrix for Majorana neutrinos can be diagonalized by a unitary matrix $V$ as
\begin{equation}
M_{\nu}=VM_{\nu}^{diag}V^{T} \label{eq1}
\end{equation}
where
\begin{center}
$M_{\nu}^{diag}$ = $\left(
\begin{array}{ccc}
m_{1} & 0 & 0 \\  0& m_{2} & 0 \\ 0& 0 & m_{3}
\end{array}
\right)$.
\end{center}
The  matrix $M_{\nu}$ can be parametrized in terms of the three neutrino masses ($m_1, m_2, m_3$), the three neutrino mixing
angles ($\theta _{12}$, $\theta _{23}$ and $\theta _{13}$, the solar, atmospheric and the reactor neutrino mixing angles, respectively) and the Dirac-type CP-violating phase $\delta$. The two additional phases ($\alpha$, $\beta$) appear for Majorana neutrinos. At present, two possible mass orderings are allowed for neutrinos: normal mass spectrum (NS) with $m_3>m_2>m_1$ or inverted mass spectrum (IS) with $m_2>m_1>m_3$.
We write the matrix $V$ as
\begin{equation}
V = UP \label{eq2}
\end{equation}
where~\cite{fogli}
\begin{equation}
U= \left(
\begin{array}{ccc}
c_{12}c_{13} & s_{12}c_{13} & s_{13}e^{-i\delta} \\
-s_{12}c_{23}-c_{12}s_{23}s_{13}e^{i\delta} &
c_{12}c_{23}-s_{12}s_{23}s_{13}e^{i\delta} & s_{23}c_{13} \\
s_{12}s_{23}-c_{12}c_{23}s_{13}e^{i\delta} &
-c_{12}s_{23}-s_{12}c_{23}s_{13}e^{i\delta} & c_{23}c_{13}
\end{array}
\right) \label{eq3}
\end{equation} with $s_{ij}=\sin\theta_{ij}$ and $c_{ij}=\cos\theta_{ij}$ and
\begin{center}
$P = \left(
\begin{array}{ccc}
1 & 0 & 0 \\ 0 & e^{i\alpha} & 0 \\ 0 & 0 & e^{i(\beta+\delta)}
\end{array}
\right)$
\end{center} 
is the diagonal phase matrix with the Majorana-type CP-violating phases $\alpha$, $\beta$ and the Dirac-type CP- violating phase $\delta$. The matrix $V$ is the neutrino mixing matrix. Using Eq. (\ref{eq1}) and Eq. (\ref{eq2}), the neutrino mass matrix can be written as
\begin{equation}
M_{\nu}=U P M_{\nu}^{diag}P^{T}U^{T}. \label{eq4}
\end{equation}
The CP-violation in neutrino oscillation experiments can be described through a rephasing invariant quantity, $J_{CP}$~\cite{jarlskog} with
$J_{CP}=Im(U_{e1}U_{\mu2}U_{e2}^*U_{\mu1}^*)$. In the above parameterization, $J_{CP}$ is given by
\begin{equation}
J_{CP} = s_{12}s_{23}s_{13}c_{12}c_{23}c_{13}^2 \sin \delta.
\end{equation}
The two texture zeros at $(p,q)$ and $(r,s)$ positions in the neutrino mass matrix give two complex equations viz.
\begin{equation}
m_1 U_{p1}U_{q1} + e^{2i\alpha}m_2 U_{p2}U_{q2} + e^{2i(\beta + \delta)}m_3 U_{p3}U_{q3} = 0 \label{eq6}
\end{equation}
and
\begin{equation}
m_1 U_{r1}U_{s1} + e^{2i\alpha}m_2 U_{r2}U_{s2} + e^{2i(\beta + \delta)}m_3 U_{r3}U_{s3} = 0 \label{eq7}
\end{equation}
where $p, q, r$ and $s$ can take the values $e$, $\mu$ and $\tau$. Solving the above two equations [Eq. (\ref{eq6}) and Eq. (\ref{eq7})] simultaneously, we obtain 
\begin{equation}
\frac{m_1}{m_2}e^{-2i\alpha }=\frac{U_{r2}U_{s2}U_{p3}U_{q3}-U_{p2}U_{q2}U_{r3}U_{s3}}{U_{p1}U_{q1}U_{r3}U_{s3}-U_{p3}U_{q3}U_{r1}U_{s1}} \label{eq8}
\end{equation}
and
\begin{equation}
\frac{m_1}{m_3}e^{-2i\beta }=\frac{U_{r3}U_{s3}U_{p2}U_{q2}-U_{p3}U_{q3}U_{r2}U_{s2}}{U_{p1}U_{q1}U_{r2}U_{s2}-U_{p2}U_{q2}U_{r1}U_{s1}}e^{2i\delta} \ . \label{eq9}
\end{equation}
The magnitudes of the two mass ratios in Eqs. (\ref{eq8}) and (\ref{eq9}), are denoted by
\begin{equation}
\eta=\left|\frac{m_1}{m_2}e^{-2i\alpha }\right|, \ \ \ \ \ \ 
\rho=\left|\frac{m_1}{m_3}e^{-2i\beta }\right| .\label{eq10}
\end{equation}
The CP-violating Majorana phases $\alpha$ and $\beta$ are given by
\begin{align}
\alpha & =-\frac{1}{2}\textrm{Arg}\left(\frac{U_{r2}U_{s2}U_{p3}U_{q3}-U_{p2}U_{q2}U_{r3}U_{s3}}{U_{p1}U_{q1}U_{r3}U_{s3}-U_{p3}U_{q3}U_{r1}U_{s1}}\right), \\
\beta & =-\frac{1}{2}\textrm{Arg}\left(\frac{U_{r3}U_{s3}U_{p2}U_{q2}-U_{p3}U_{q3}U_{r2}U_{s2}}{U_{p1}U_{q1}U_{r2}U_{s2}-U_{p2}U_{q2}U_{r1}U_{s1}}e^{2i\delta}\right).
\end{align}
The two mass ratios ($\eta, \rho$) and the two Majorana-type CP-violating phases ($\alpha, \beta$) are obtained in terms of three neutrino mixing angles ($\theta_{12}, \theta_{13}, \theta_{23}$) and the Dirac-type CP-violating phase $\delta$.\\ The two mass ratios can be further used to obtain two values of $m_1$ viz.
\begin{equation}
m_{1}=\eta \sqrt{\frac{ \Delta
m_{21}^{2}}{1-\eta ^{2}}} \ , \ \ 
m_{1}=\rho \sqrt{\frac{\Delta m_{21}^{2}+
|\Delta m_{23}^{2}|}{ 1-\rho^{2}}} \label{eq13}
\end{equation}
where ($\Delta m_{ij}^2 \equiv m_i^2 - m_j^2$). The above two values of $m_1$ contain the constraints of two texture zeros in $M_\nu$ through the two mass ratios $\eta$ and $\rho$. The simultaneous existence of two texture zeros in $M_\nu$ requires these two values of $m_1$ to be equal.\\
In the case of two texture zeros, there exists a permutation symmetry between different patterns. This corresponds to the permutation of the 2-3 rows and 2-3 columns of $M_\nu$. The corresponding permutation matrix is given by
\begin{equation}
P_{23} = \left(
\begin{array}{ccc}
1&0&0\\
0&0&1\\
0&1&0\\
\end{array}
\right). \label{eq14}
\end{equation}
This leads to the following relations between the parameters of the classes related by the permutation symmetry: 
\begin{equation}
\theta_{12}^{X} = \theta_{12}^{Y}, \ \theta_{13}^{X} = \theta_{13}^{Y}, \ \theta_{23}^{X} = \frac{\pi}{2}-\theta_{23}^{Y}, \ \delta^{X} = \delta^{Y} - \pi \ 
\end{equation}
where $X$ and $Y$ denote classes related by the permutation symmetry.
The classes related by the 2-3 permutation symmetry are
\begin{equation}
A_1 \leftrightarrow A_2, \ B_1 \leftrightarrow B_2, \ B_3 \leftrightarrow B_4
\end{equation}
whereas the class $C$ transforms unto itself.
\section{Numerical Analysis}
The effective Majorana mass $|M_{ee}|$ which determines the rate of neutrinoless double beta (NDB) decay is given by 
\begin{equation}
|M_{ee}|= |m_1c_{12}^2c_{13}^2+ m_2s_{12}^2c_{13}^2 e^{2i\alpha}+ m_3s_{13}^2e^{2i\beta}|.
\end{equation}
The observation of NDB decay would imply lepton number violation and the Majorana nature of neutrinos. For recent reviews on NDB decay see~\cite{ndbdecay,ndbrodejohann}. A large number of projects such as CUORICINO~\cite{cuoricino}, CUORE~\cite{cuore}, GERDA~\cite{gerda}, MAJORANA~\cite{majorana}, SuperNEMO~\cite{supernemo}, EXO~\cite{exo}, GENIUS~\cite{genius} aim to achieve a sensitivity up to 0.01 eV for $|M_{ee}|$. In the present work, we take the upper limit on $|M_{ee}|$ to be 0.5 eV~\cite{ndbrodejohann}. In addition, cosmological observations put an upper bound on the sum of light neutrino masses
\begin{equation}
\Sigma = \sum_{i=1}^3 m_i \ .
\end{equation} 
Data from Planck satellite~\cite{planck} combined with other cosmological data limit $\Sigma < 0.23$ at $95\%$ confidence level (CL). However, these bounds are strongly dependent on model details and the data set used. In the numerical analysis, we take the conservative upper limit $\Sigma < 1$ eV. The recent experimental results on neutrino oscillation parameters at 1, 2 and 3$\sigma$ CL~\cite{valledata} are given in Table~\ref{tab2}.
\begin{table}[t!]
\begin{center}
\begin{tabular}{|c|c|}
\hline Parameter & Mean $^{(+1 \sigma, +2 \sigma, +3 \sigma)}_{(-1 \sigma, -2 \sigma, -3 \sigma)}$ \\
\hline $\Delta m_{21}^{2} [10^{-5}eV^{2}]$ & $7.62_{(-0.19,-0.35,-0.5)}^{(+0.19,+0.39,+0.58)}$ \\ 
\hline $\Delta m_{31}^{2} [10^{-3}eV^{2}]$ & $2.55_{(-0.09,-0.19,-0.24)}^{(+0.06,+0.13,+0.19)}$, \\&
$(-2.43_{(-0.07,-0.15,-0.21)}^{(+0.09,+0.19,+0.24)})$ \\ 
\hline $\sin^2 \theta_{12}$ & $0.32_{(-0.017,-0.03,-0.05)}^{(+0.016,+0.03,+0.05)}$ \\ 
\hline $\sin^2 \theta_{23}$ & $0.613_{(-0.04,-0.233,-0.25)}^{(+0.022,+0.047,+0.067)}$, \\& $(0.60_{(-0.031,-0.210,-0.230)}^{(+0.026,+0.05,+0.07)})$ \\ 
\hline $\sin^2 \theta_{13}$ & $0.0246_{(-0.0029,-0.0054,-0.0084)}^{(+0.0028,+0.0056,+0.0076)}$,\\& $(0.0250_{(-0.0027,-0.005,-0.008)}^{(+0.0026,+0.005,+0.008)})$ \\ 
\hline 
\end{tabular}
\caption{\label{tab2}Current neutrino oscillation parameters from global fits~\cite{valledata}. The upper (lower) row corresponds to normal (inverted) spectrum, with $\Delta m^2_{31} > 0$ ($\Delta m^2_{31} < 0$).}
\end{center}
\end{table}
In the numerical analysis, first, we use the experimental input of the two mass squared differences ($\Delta m_{21}^2, \Delta m_{23}^2$) along with the constraints of two texture zeros and large $|M_{ee}|$ to obtain predictions for mixing angles. We vary the two mass squared differences randomly within the 3$\sigma$ allowed ranges but keep the neutrino mixing angles free and vary them between $0^\circ$ and $90^\circ$. The Dirac phase is varied from $0^\circ$ to $360^\circ$ and the constraint of a large $|M_{ee}| > 0.08$ eV is imposed. The two values of $m_1$ obtained in Eq. (\ref{eq13}) should be equal within the errors of the oscillation parameters for the simultaneous existence of two texture zeros in $M_\nu$. We carry out this analysis for classes $B_1, B_2, B_3$ and $B_4$.\\
\begin{figure}[t!]
\begin{center}
{\epsfig{file=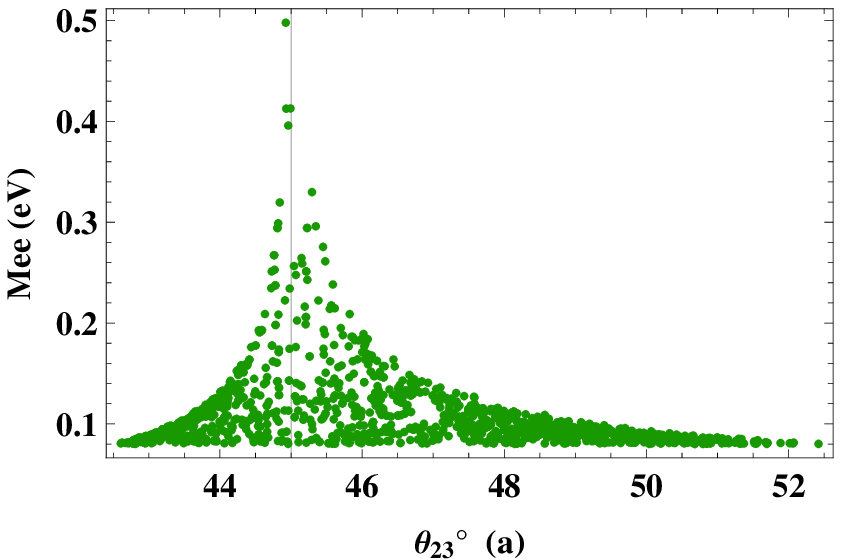, width=5.0cm, height=4.0cm} 
\epsfig{file=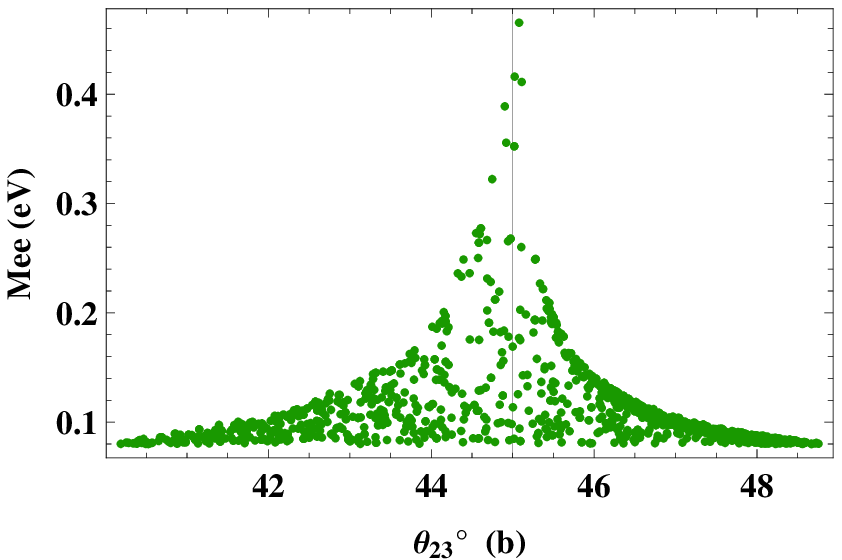, width=5.0cm, height=4.0cm}\\ 
\epsfig{file=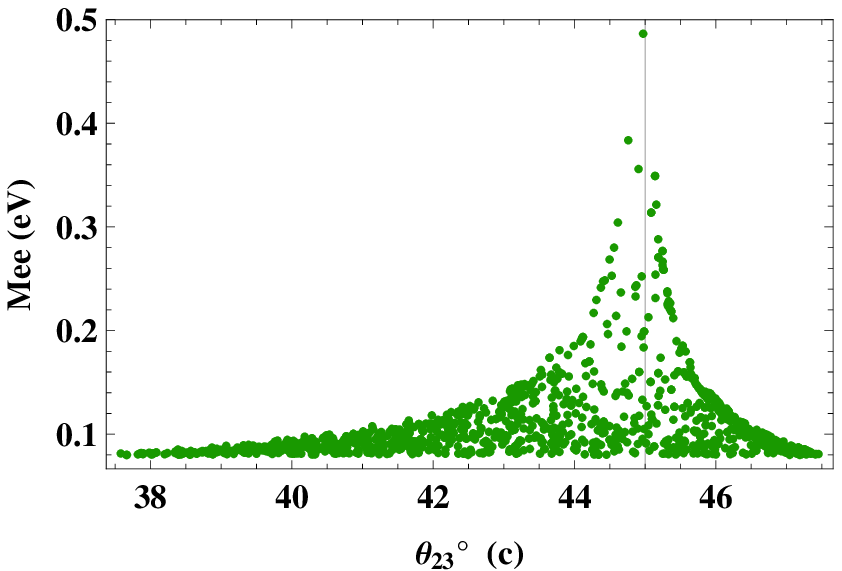, width=5.0cm, height=4.0cm} 
\epsfig{file=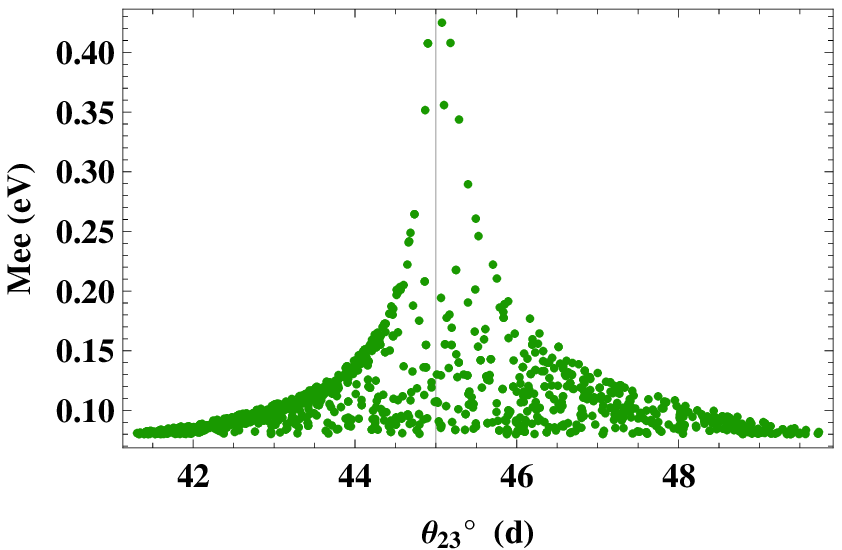, width=5.0cm, height=4.0cm} 
}
\caption{\label{fig1}Correlation plots for classes $B_1$(NH)(a), $B_1$(IH)(b), $B_2$(NH)(c) and $B_2$(IH)(d). Here the mixing angles are varied between $0^\circ$ and $90^\circ$.}
\end{center}
\end{figure}
\begin{figure}[t!]
\begin{center}
{\epsfig{file=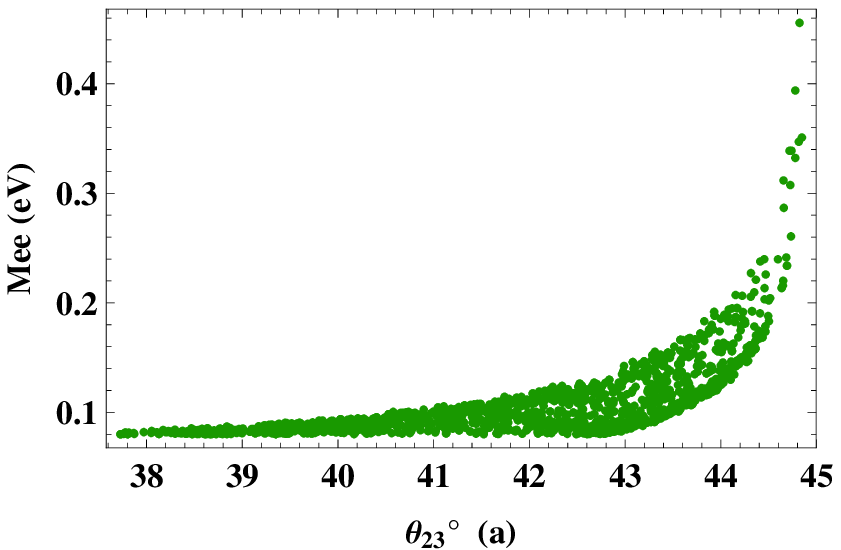, width=5.0cm, height=4.0cm} 
\epsfig{file=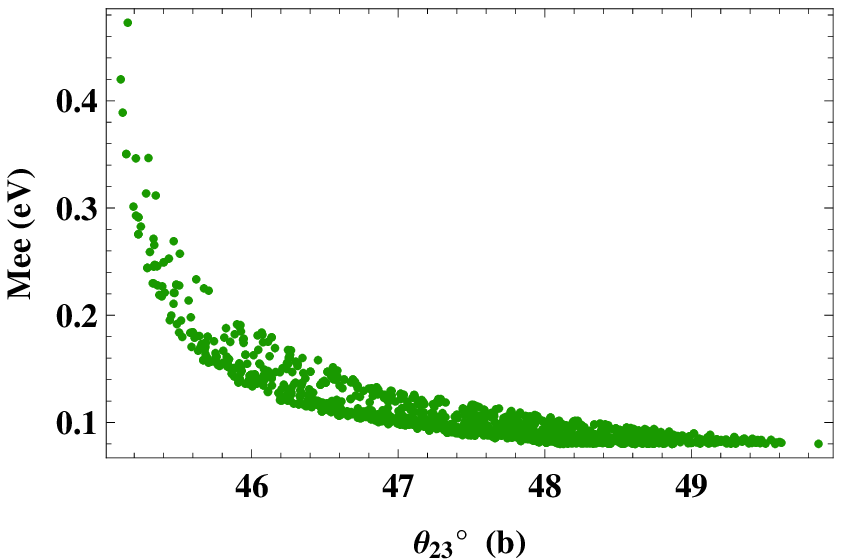, width=5.0cm, height=4.0cm}\\ 
\epsfig{file=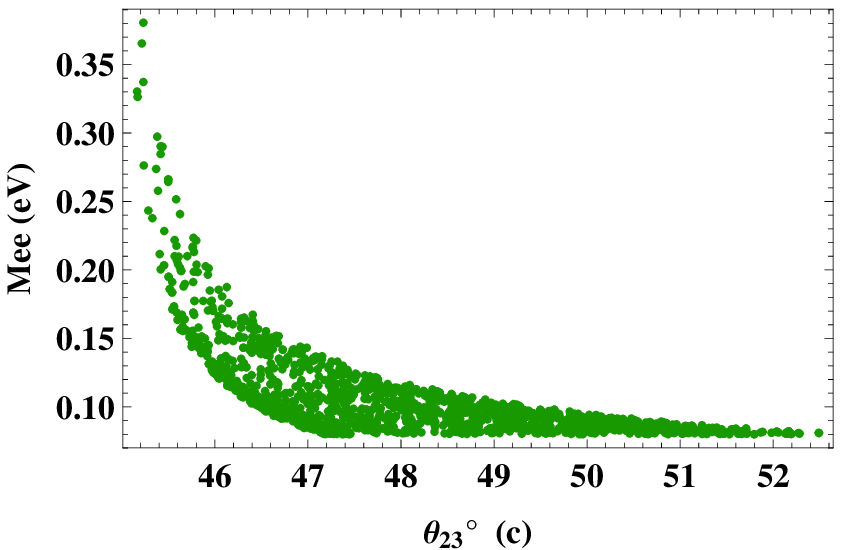, width=5.0cm, height=4.0cm} 
\epsfig{file=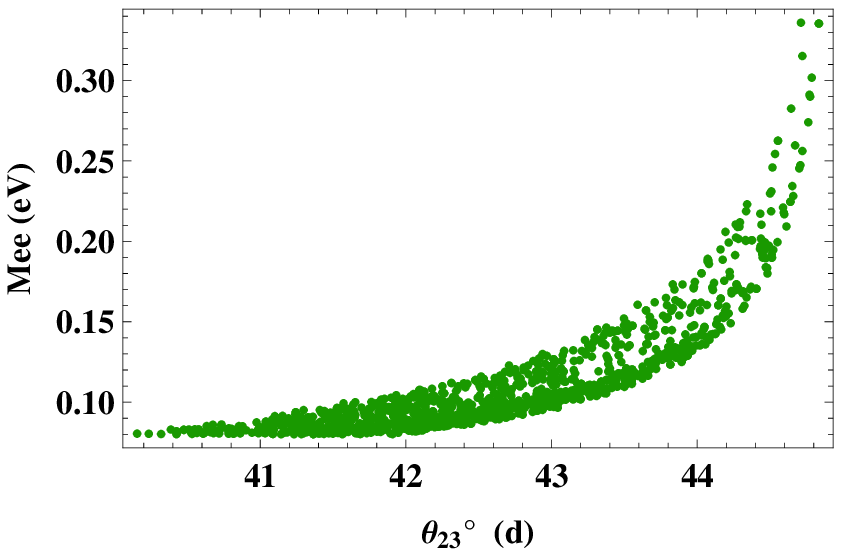, width=5.0cm, height=4.0cm} 
}
\caption{\label{fig2}Correlation plots for classes $B_3$(NH)(a), $B_3$(IH)(b), $B_4$(NH)(c) and $B_4$(IH)(d). Here the mixing angles are varied between $0^\circ$ and $90^\circ$.}
\end{center}
\end{figure}
\begin{figure}[t!]
\begin{center}
{\epsfig{file=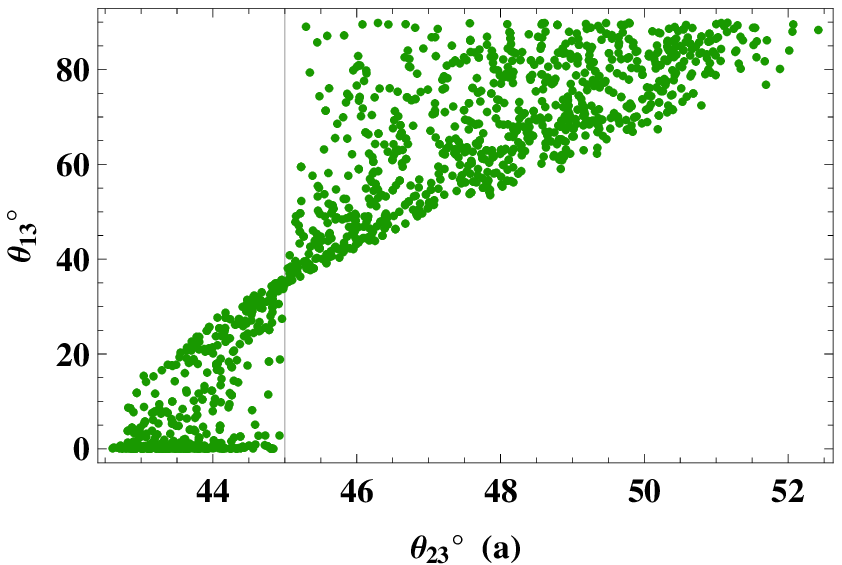, width=5.0cm, height=4.0cm} 
\epsfig{file=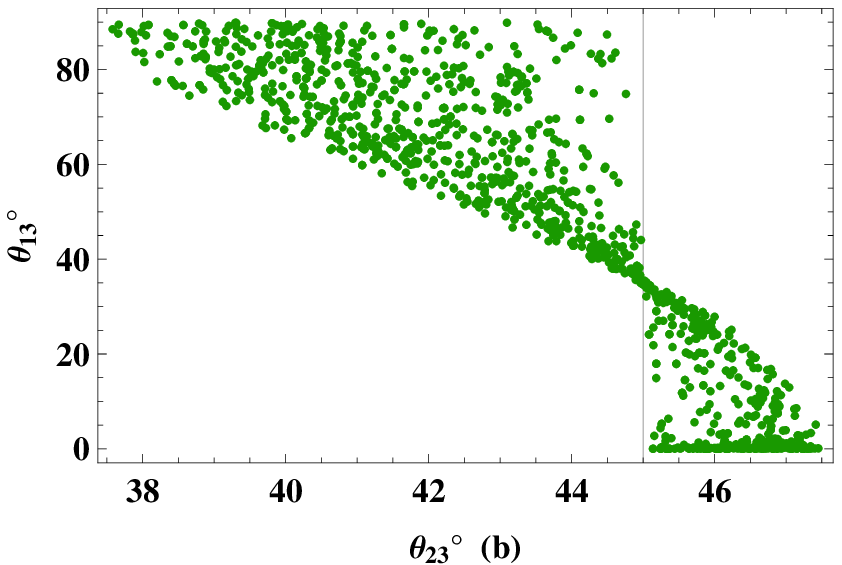, width=5.0cm, height=4.0cm}\\ 
\epsfig{file=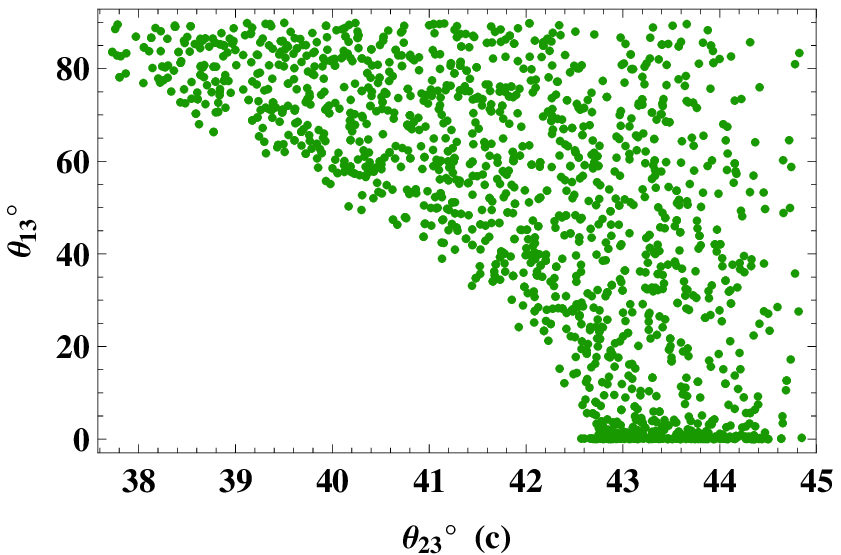, width=5.0cm, height=4.0cm} 
\epsfig{file=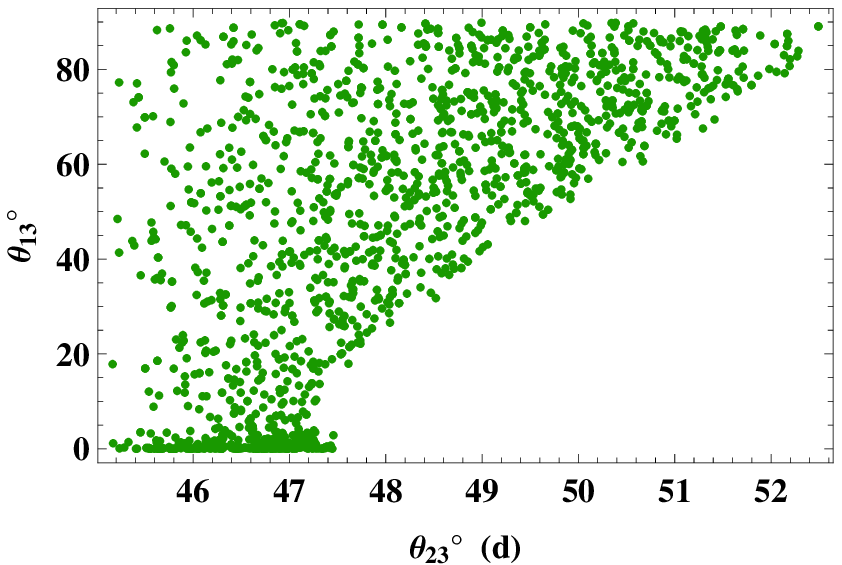, width=5.0cm, height=4.0cm} 
}
\caption{\label{fig3}Correlation plots (NH) for classes $B_1$(a), $B_2$(b) and $B_3$(c), $B_4$(d) depicting the 2-3 interchange symmetry. Here the mixing angles are varied between $0^\circ$ and $90^\circ$.}
\end{center}
\end{figure}
It is found that all these classes predict a near maximal atmospheric neutrino mixing angle while the other two mixing angles remain unconstrained. The atmospheric neutrino mixing angle remains near maximal irrespective of the values of solar and reactor neutrino mixing angles. It can be inferred from Fig.~\ref{fig1} and Fig.~\ref{fig2} that for all these classes, $\theta_{23}$ moves towards $45^\circ$ with increasing $|M_{ee}|$. Fig~\ref{fig3} depicts the 2-3 interchange symmetry between classes $B_1 \leftrightarrow B_2$ and $ B_3 \leftrightarrow B_4$. It can be seen from Fig.~\ref{fig2} and Fig.~\ref{fig3} that for classes $B_3$ and $B_4$ the quadrant of $\theta_{23}$ is already decided without the experimental input of the mixing angles.\\ Grimus et al.~\cite{grimus} have shown that only for classes $B_3$ and $B_4$ a near maximal $\theta_{23}$ is predicted in the limit of a quasi degenerate (QD) spectrum. In comparison, our assumption of a large $|M_{ee}|$ apart from ensuring a QD spectrum puts additional constraints on mixing angles and CP-violating phases which leads to a near maximal $\theta_{23}$ for all the four classes ($B_1$, $B_2$, $B_3$ and $B_4$). Comparing the correlation plots in Fig.~\ref{fig1} and Fig.~\ref{fig3}(a, b) with those presented in Ref.~\cite{ourzm}, it is clear that classes $B_1, B_2$ of two texture zeros have the same phenomenological predictions (except for neutrino mass spectrum) as those for classes $B_6, B_5$ of two vanishing cofactors, respectively. This similarity is not coincidental and, in fact, it has been shown in Ref.~\cite{dualmodels} that for any model with some homogeneous relationship between the elements of the effective neutrino mass matrix with one mass spectrum, there are similar predictions for the oscillation parameters and the Majorana phases as for models with the same relationship among cofactors of the effective neutrino mass matrix with opposite neutrino mass spectrum.\\
\begin{figure}[t!]
\begin{center}
{\epsfig{file=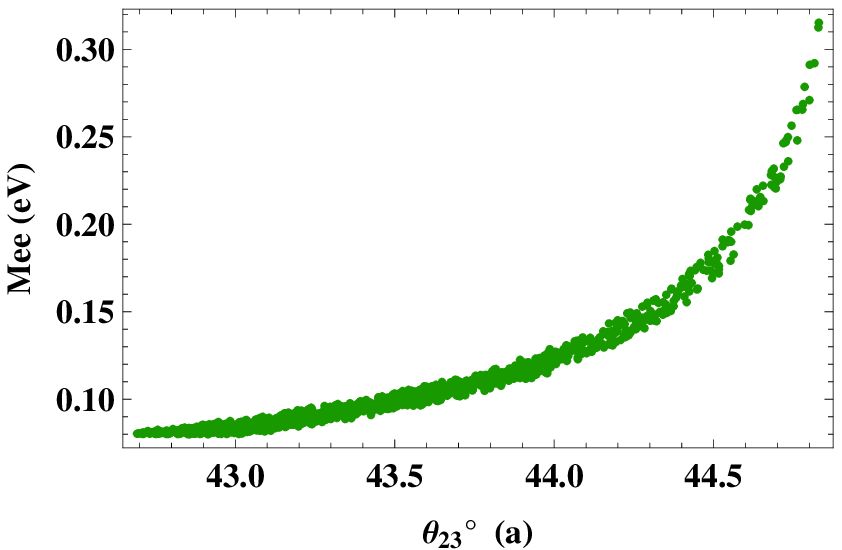, width=5.0cm, height=4.0cm} 
\epsfig{file=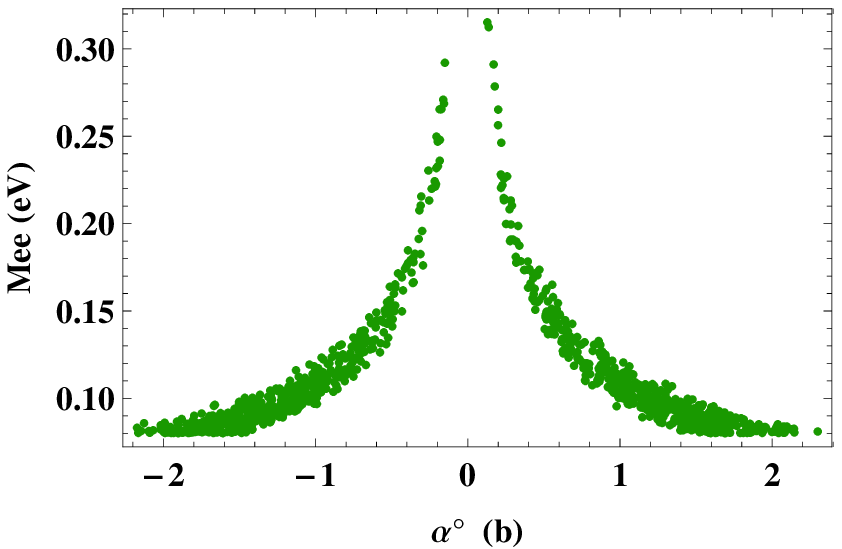, width=5.0cm, height=4.0cm}\\ 
\epsfig{file=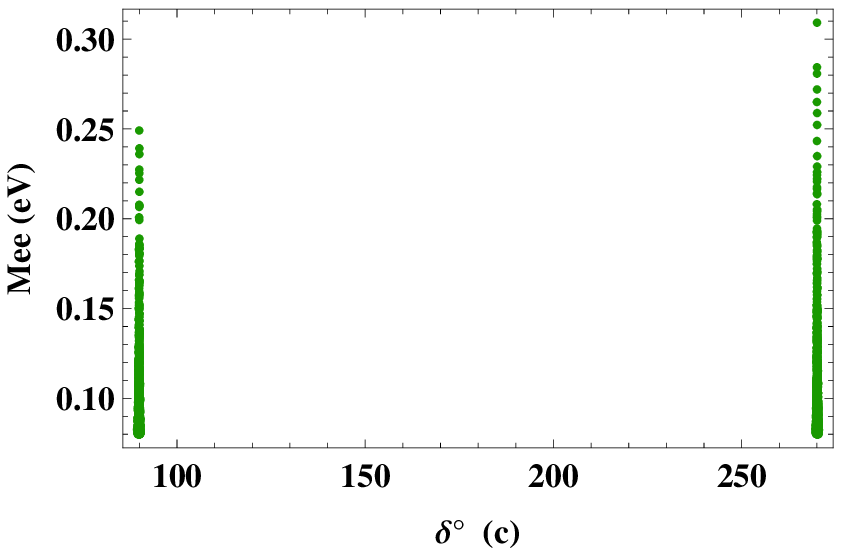, width=5.0cm, height=4.0cm} 
\epsfig{file=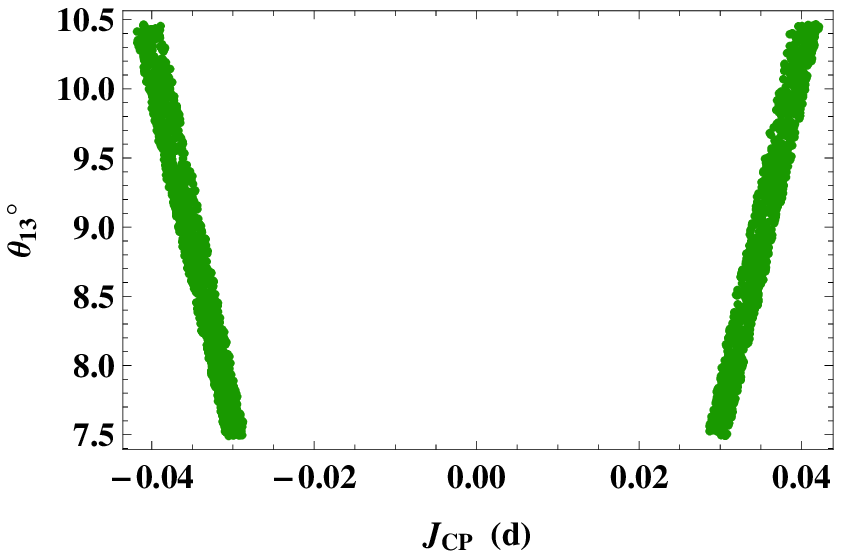, width=5.0cm, height=4.0cm} 
}
\caption{\label{fig4}Correlation plots for class $B_1$(NH)(a, b) and $B_1$(IH)(c, d).}
\end{center}
\end{figure}
\begin{figure}[t!]
\begin{center}
{\epsfig{file=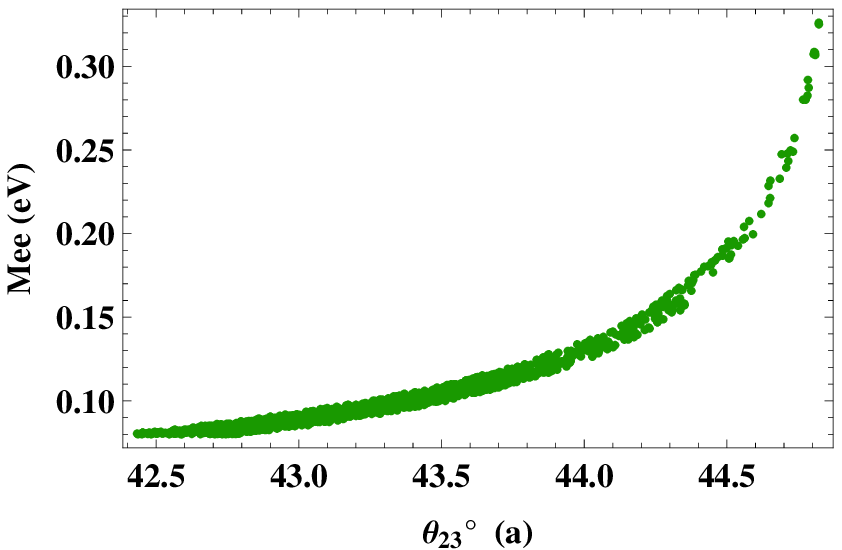, width=5.0cm, height=4.0cm} 
\epsfig{file=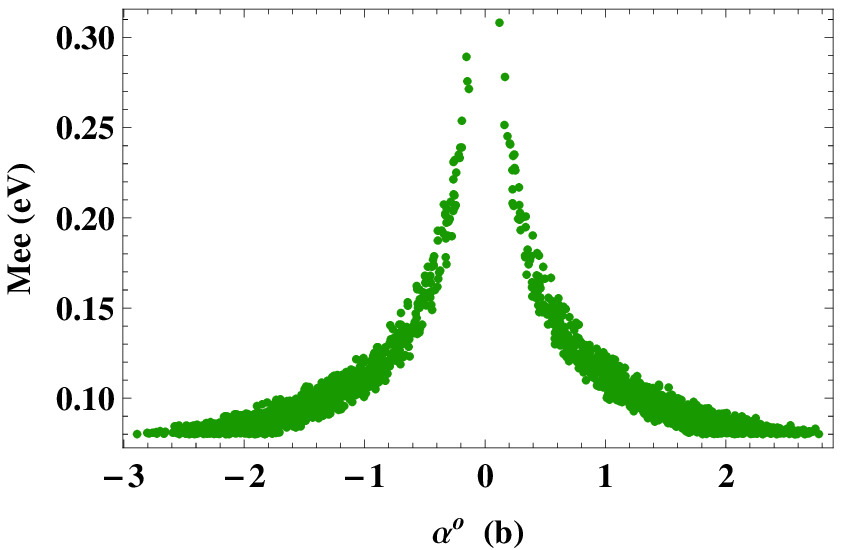, width=5.0cm, height=4.0cm}\\ 
\epsfig{file=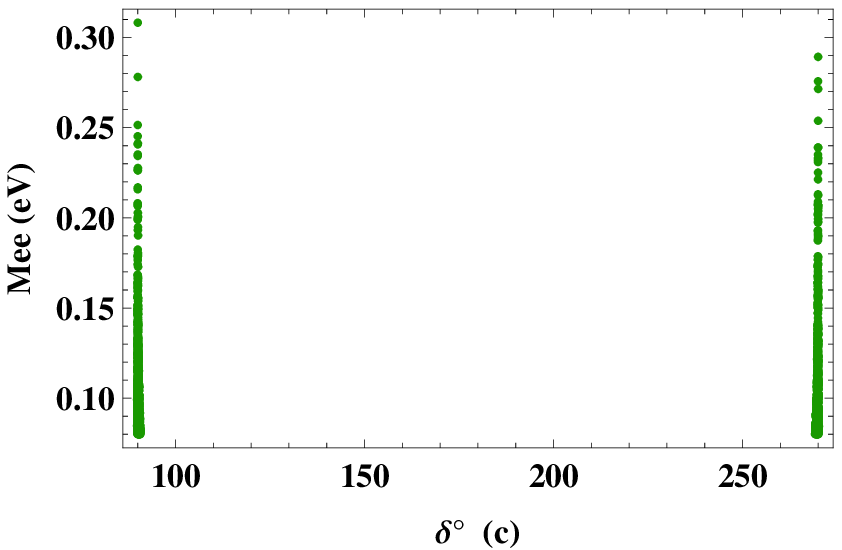, width=5.0cm, height=4.0cm} 
\epsfig{file=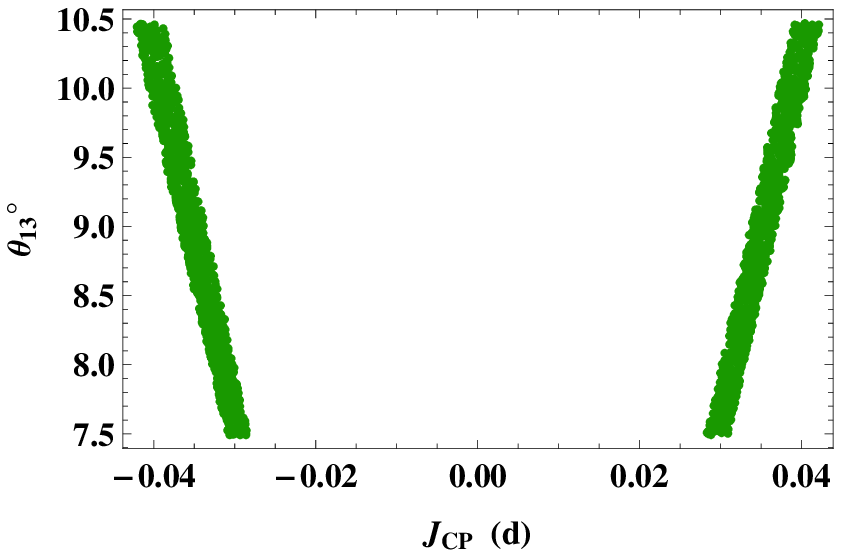, width=5.0cm, height=4.0cm} 
}
\caption{\label{fig5}Correlation plots for class $B_3$(NH)(a, b) and $B_3$(IH)(c, d).}
\end{center}
\end{figure}
In the second step, we also take into account the experimental input of mixing angles along with the experimental input of the two mass squared differences. With these additional inputs, the phenomenological predictions of the four classes $B_1$, $B_2$, $B_3$ and $B_4$ substantially overlap (see Fig.~\ref{fig4} and Fig.~\ref{fig5}) and a precise determination of $\delta$ and the determination of neutrino mass spectrum is crucial to pin down one of the four classes \cite{ourdegeneracies,xingtz,peinado}. The correlation plots for classes $B_1$ and $B_3$ are shown in Fig.~\ref{fig4} and Fig.~\ref{fig5}, respectively. One can see from Fig.~\ref{fig4}(d) and Fig.~\ref{fig5}(d) that $J_{CP}$ cannot vanish for these classes which implies that these classes are necessarily CP-violating. For all the above four classes, the Dirac-type CP-violating phase is predicted to be very close to $90^\circ$ or $270^\circ$.
\begin{table}[t!]
\begin{tiny}
\begin{center}
\begin{tabular}{|c|c|c|c|c|c|}
\hline  Class & MS & $\delta$ & $\theta_{23}$& $M_\nu$ & LO \\
\hline $B_1$ & NS & $90.65^\circ$ & $43.64^\circ$ & $\left(
\begin{array}{ccc}
0.10479 + 0.00127 i & 0.00018 + 0.00258i & 6.486 \times 10^{-9} - 9.31 \times 10^{-10} i\\
0.00018 + 0.00258 i & 0.0 - 3.260 \times 10^{-9} i & -0.11001 - 0.00287 i\\
 6.486 \times 10^{-9} - 9.31 \times 10^{-10}i & -0.11001 - 0.00287 i & -0.01105 - 0.00024\\
\end{array}
\right)$  &  $\left(
\begin{array}{ccc}
0.1 & \lambda^2 & 0\\
\lambda^2 & 0 & 0.1\\
0 & 0.1 & \lambda\\
\end{array}
\right)$ \\ \cline{2-5} & IS & $ 89.91^\circ$& $ 47.87^\circ$ &$\left(
\begin{array}{ccc}
0.08919 + 0.00151 i & 0.00021 + 0.00317 i & -4.409 \times 10^{-9} + 0.0 i\\
0.00021 + 0.00317 i & 0.0 - 5.588 \times 10^{-9}i & -0.08200 -0.00293 i\\
-4.409 \times 10^{-9} + 0.0 i & -0.08200 -0.00293 i & 0.01469 + 0.00045 i\\
\end{array}
\right)$ & \\ 
\hline $B_3$ & NS & $ 89.46^\circ$ & $ 43.77^\circ$ & $\left(
\begin{array}{ccc}
0.11722 - 0.00104 i & 0.0 - 4.66 \times 10^{-10} i & -0.00019 + 0.00222 i\\
0.0 - 4.66 \times 10^{-10} i  & 7.451 \times 10^{-9} 9.31 \times 10^{-10} i & -0.12195 + 0.00239 i\\
-0.00019 + 0.00222 i & -0.12195 + 0.00239 i & -0.00997 + 0.00024 i\\
\end{array}
\right)$ & $\left(
\begin{array}{ccc}
0.1 & 0 & \lambda^2\\
0 & 0 & 0.1\\
\lambda^2 & 0.1 & \lambda\\
\end{array}
\right)$ \\ \cline{2-5} & IS  & $ 90.20^\circ$ & $ 47.43^\circ$ &$\left(
\begin{array}{ccc}
0.09327 + 0.00169 i & 0.0 - 1.16 \times 10^{-10} i & -0.00016 - 0.00327 i\\
0.0 - 1.16 \times 10^{-10} i & 3.725 \times 10^{-9} + 9.31 \times 10^{-10} i & -0.08648 -0.00325 i\\
-0.00016 - 0.00327 i & -0.08648 -0.00325 i & 0.01392 + 0.00059 i\\
\end{array}
\right)$ & \\ 
\hline 
\end{tabular}
\end{center}
\end{tiny}
\caption{\label{tab3}Numerically estimated neutrino mass matrices for classes $B_1$ and $B_3$. MS denotes the neutrino mass spectrum and LO in the last column denotes the estimated magnitudes of the mass matrix elements at leading order with $\lambda \sim 0.01$. }
\end{table}     
The numerically estimated mass matrices for classes $B_1$ and $B_3$ are given in Table~\ref{tab3}. These mass matrices are obtained for the best fit values of $\Delta m_{21}^2, \Delta m_{23}^2, \theta_{12}$ and $\theta_{13}$. The numerical mass matrices for $B_2$ and $B_4$ patterns can be obtained from Table~\ref{tab3} with the operation of $P_{23}$ [Eq. (\ref{eq14})] on the mass matrices of patterns $B_1$ and $B_3$, respectively.    
\section{Symmetry Realization}
In this section, we show how the texture zero structures discussed in this work can be obtained using discrete Abelian family symmetries. General guidelines for the symmetry realization of texture zeros in both the quark and the lepton mass matrices have been enunciated in Ref.~\cite{grimussym} which outlines procedures for enforcing texture zeros at any place in the fermion mass matrices by imposing discrete Abelian family symmetries. To obtain the desired texture  structures for class $B$, we use the framework of type-I+II seesaw mechanism~\cite{seesaw1,seesaw2}.\\
The type-I seesaw~\cite{seesaw1} contribution to the effective neutrino mass matrix is given by
\begin{equation}
M_\nu^I \approx -M_D M_R^{-1} M_D^T
\end{equation}
where $M_D$ and $M_R$ are the Dirac and the right-handed neutrino mass matrices, respectively. For the type-I seesaw mechanism to work, the standard model (SM) is extended by adding three right-handed neutrinos ($\nu_{eR}, \nu_{\mu R}, \nu_{\tau R}$). For classes $B_3$ and $B_4$, we will also need one $SU(2)_L$ singlet scalar ($\chi$). For the type-II seesaw \cite{seesaw2} contribution to the effective neutrino mass matrix, we need a scalar $SU(2)_L$ triplet Higgs ($\triangle$). The effective neutrino mass matrix containing both type-I+II seesaw contributions is given by
\begin{equation}
M_\nu \approx M_\nu^{II} + M_\nu^I = M_L - M_D M_R^{-1} M_D^T
\end{equation}
where $M_L$ denotes the type-II seesaw contribution.
The symmetry realization of all the presently allowed classes of two texture zeros in $M_\nu$ in the flavor basis has been presented in~\cite{grimustz, xingtz} in the context of type-II seesaw mechanism and in Ref.~\cite{ourtzsym} considering both type-I+II seesaw contributions and using a minimal cyclic symmetry group. Classes $B_1$ and $B_2$  have also been realized in Ref.~\cite{joshipuratz} using $A_4$ or its $Z_3$ subgroup in the context of type-I+II seesaw mechanism. Also, classes $B_3$ and $B_4$ have been obtained in Ref.~\cite{schmidt} by softly breaking the $L_\mu - L_\tau$ symmetry.\\
For the symmetry realization, we use the discrete cyclic group $Z_3$. For class $B_1$, we assume the following transformation properties of the leptonic fields under the action of $Z_3$ symmetry:
\begin{align}
 D_{eL}&\rightarrow \omega D_{eL},& e_R & \rightarrow \omega e_R,& \nu_{e R}&\rightarrow \omega \nu_{e R},&  \nonumber \\ D_{\mu L}& \rightarrow \omega^2 D_{\mu_L},& \mu_R & \rightarrow \omega^2 \mu_R ,&  \nu_{\mu R} & \rightarrow  \omega^2 \nu_{\mu R},&  \\ D_{\tau L} & \rightarrow D_{\tau L},& \tau_R & \rightarrow \tau_R ,&  \nu_{\tau R} & \rightarrow  \nu_{\tau R},& \nonumber
\end{align}
where $\omega$ = $e^{i 2 \pi/3}$. $D_{lL}$ $(l = e, \mu, \tau)$ denotes $SU(2)_L$ doublets and $l_R, \nu_{lR}$ denote the right-handed $SU(2)_L$ singlet charged lepton and neutrino fields, respectively. According to the above transformations of the leptonic fields, the bilinears $\overline{D}_{l L} l_R$, $\overline{D}_{l L}\nu_{l R}$ and $\nu_{l R}^T C^{-1} \nu_{l R}$ relevant for $M_l$, $M_D$ and $M_R$ respectively, transform as
\begin{small} 
\begin{equation}
\overline{D}_{lL} l_R \sim \overline{D}_{l L}\nu_{l R} \sim \left(
\begin{array}{ccc}
1& \omega &\omega^2 \\
\omega^2 &1 &\omega \\
\omega& \omega^2&1
\end{array}
\right), \ \ \ \ \ \ \ \nu_{l R}^T C^{-1} \nu_{l R} \sim \left(
\begin{array}{ccc}
\omega^2& 1 &\omega \\
1 &\omega &\omega^2 \\
\omega& \omega^2&1
\end{array}
\right).
\end{equation}
\end{small}
The SM Higgs doublet is assumed to remain invariant under $Z_3$ leading to diagonal $M_l$, $M_D$. $M_R$ has the following form:
\begin{small} 
\begin{equation}
M_R = \left(
\begin{array}{ccc}
0&C&0\\
C&0&0\\
0&0&D\\
\end{array}
\right).
\end{equation}
\end{small}
This leads to the following type-I seesaw contribution to the effective neutrino mass matrix:
\begin{small} 
\begin{equation}
M_\nu^I = \left(
\begin{array}{ccc}
0&c&0\\
c&0&0\\
0&0&d\\
\end{array}
\right).
\end{equation}
\end{small}
We add a Higgs triplet $\triangle$ which transforms as $\triangle \rightarrow \omega \triangle$ under $Z_3$, leading to the following type-II seesaw contribution to the effective neutrino mass matrix:
\begin{small} 
\begin{equation}
M_\nu^{II} = \left(
\begin{array}{ccc}
a&0&0\\
0&0&b\\
0&b&0\\
\end{array}
\right).
\end{equation}
\end{small} 
The effective neutrino mass matrix after the complete type-I+II seesaw contributions has the form of class $B_1$ viz.
\begin{small} 
\begin{equation}
M_\nu \equiv M_\nu^I + M_\nu^{II} = \left(
\begin{array}{ccc}
a&c&0\\
c&0&b\\
0&b&d\\
\end{array}
\right).
\end{equation}
\end{small} 
\begin{table}[t!]
\begin{small}
\noindent\makebox[\textwidth]{
\begin{tabular}{|c|c|c|c|c|c|c|}
\hline Class & $D_{eL}$, \ $D_{\mu L}$, \ $D_{\tau L}$ & $e_R$, \ $\mu_R$, \ $\tau_R$ & $\nu_{e R}$, \ $\nu_{\mu R}$, \ $\nu_{\tau R}$ & $\phi$ & $\triangle$ & $\chi$ \\
\hline $B_1$ & $\omega$, \ $\omega^2$, \ $1$ & $\omega$, \ $\omega^2$, \ $1$  &  $\omega$, \ $\omega^2$, \ $1$  & $1$ & $\omega$ & - \\
\hline $B_2$ & $\omega^2$, \ $1$, \ $\omega$   & $\omega^2$, \ $1$, \ $\omega$ &  $\omega^2$, \ $1$, \ $\omega$ & $1$ & $\omega^2$ & - \\
\hline $B_3$ & $1$, \ $\omega^2$, \ $\omega$ & $1$, \ $\omega^2$, \ $\omega$ & $\omega$, \ $1$, \ $1$  & $1$ & $1$ & $\omega^2$\\
\hline $B_4$ & $1$, \ $\omega$, \ $\omega^2$ & $1$, \ $\omega$, \ $\omega^2$ & $\omega$, \ $1$, \ $1$  & $1$ & $1$ & $\omega^2$\\
\hline
\end{tabular}}
\end{small}
\caption{\label{tab4}The transformation properties of lepton and scalar fields under $Z_3$ for classes $B_1$, $B_2$, $B_3$ and $B_4$.}
\end{table}
\begin{table}[t!]
\begin{small}
\noindent\makebox[\textwidth]{
\begin{tabular}{|c|c|c|c|c|c|}
\hline Class & $M_D$ & $M_R$ & $M_\nu^I$ & $M_\nu^{II}$ & $M_\nu$ \\
\hline $B_1$ & $\left(
\begin{array}{ccc}
x & 0 & 0 \\  0 & y & 0 \\ 0 & 0 & z
\end{array}
\right)$ & $\left(
\begin{array}{ccc}
0 & C & 0 \\  C & 0 & 0 \\ 0 & 0 & D
\end{array}
\right)$ & $\left(
\begin{array}{ccc}
0 & c & 0 \\  c & 0 & 0 \\ 0 & 0 & d
\end{array}
\right)$ & $\left(
\begin{array}{ccc}
a & 0 & 0 \\  0 & 0 & b \\ 0 & b & 0
\end{array}
\right)$&$\left(
\begin{array}{ccc}
a & c & 0 \\  c & 0 & b \\ 0 & b & d
\end{array}
\right)$\\
\hline $B_2$ & $\left(
\begin{array}{ccc}
x & 0 & 0 \\  0 & y & 0 \\ 0 & 0 & z
\end{array}
\right)$ & $\left(
\begin{array}{ccc}
0 & 0 & C \\  0 & D &0 \\ C & 0 & 0
\end{array}
\right)$ & $\left(
\begin{array}{ccc}
0 & 0& c \\  0 & d & 0 \\ c & 0 & 0
\end{array}
\right)$  & $\left(
\begin{array}{ccc}
a & 0 & 0 \\  0 & 0 & b \\ 0 & b & 0
\end{array}
\right)$ & $\left(
\begin{array}{ccc}
a & 0 & c \\  0 & d & b \\ c & b & 0
\end{array}
\right)$ \\
\hline $B_3$ & $\left(
\begin{array}{ccc}
0 & x & y \\  0 & 0 & 0 \\ z & 0 & 0
\end{array}
\right)$ & $\left(
\begin{array}{ccc}
0 & A & B \\  A & C & D \\ B & D & E
\end{array}
\right)$  & $\left(
\begin{array}{ccc}
c & 0 & d \\  0 & 0 & 0 \\ d & 0 & e
\end{array}
\right)$ & $\left(
\begin{array}{ccc}
a & 0 & 0 \\  0 & 0 & b \\ 0 & b & 0
\end{array}
\right)$  & $\left(
\begin{array}{ccc}
a+c & 0 & d \\  0 & 0 & b \\ d & b & e
\end{array}
\right)$ \\
\hline $B_4$ & $\left(
\begin{array}{ccc}
0 & x & y \\  z & 0 & 0 \\ 0 & 0 & 0
\end{array}
\right)$ & $\left(
\begin{array}{ccc}
0 & A & B \\  A & C & D \\ B & D & E
\end{array}
\right)$  & $\left(
\begin{array}{ccc}
c & d & 0 \\  d & e & 0 \\ 0 & 0 & 0
\end{array}
\right)$ & $\left(
\begin{array}{ccc}
a & 0 & 0 \\  0 & 0 & b \\ 0 & b & 0
\end{array}
\right)$  & $\left(
\begin{array}{ccc}
a+c & d & 0 \\  d & e & b \\ 0 & b & 0
\end{array}
\right)$\\
\hline
\end{tabular}}
\end{small}
\caption{\label{tab5}Structures of $M_{D}$, $M_{R}$, type-I and type-II seesaw contributions to the effective neutrino mass matrix.}
\end{table}
In Table~\ref{tab4}, we have summarized the transformation properties of lepton and scalar fields under the action of $Z_3$ for classes $B_1$, $B_2$, $B_3$ and $B_4$.\\
For classes $B_3$ and $B_4$ we need an $SU(2)_L$ singlet scalar for obtaining the desired form of $M_R$ (see table \ref{tab5}) corresponding to these two classes. We have closely followed our earlier work \cite{ourtzsym} for the symmetry realization of classes $B_3$ and $B_4$. Also, for the symmetry realization of classes $B_1$ and $B_2$ we need to softly break the scalar potential by allowing the $Z_3$ forbidden, dimension three term $\phi^\dagger \triangle \tilde{\phi}$ \footnote{We thank E. Peinado for drawing our attention to this point.}. This term is required to obtain a non-zero and small VEV of the scalar triplet ($\triangle$)\cite{grimustz, grimusreview, matriplet}.\\
From Table~\ref{tab3}, we can see that for each class there is a hierarchy between the magnitudes of the non-zero elements of $M_\nu$. The magnitude of $(e,e)$ and $(\mu,\tau)$ elements is at least an order of magnitude greater than the rest of the non-zero elements of $M_\nu$. 
To account for the hierarchy between the non-zero elements of the neutrino mass matrices of classes $B_1$, $B_2$, $B_3$ and $B_4$, we assume the dominance of type-II seesaw contribution. As can be seen from Table~\ref{tab5}, for all the classes ($B_1$, $B_2$, $B_3$, $B_4$) considered in this work, the dominant type-II seesaw contributes to $(e,e)$ and $(\mu,\tau)$ elements of $M_\nu$ in each case, thus, accounting for the hierarchy between the non-zero elements of $M_\nu$. 
\section{Stability of Texture Zeros}
All the above classes of texture zeros are realized at the seesaw scale which poses the question whether the texture zeros realized in this work survive when the RG evolution of $M_\nu$ from the seesaw to the electroweak scale is taken into account. At one loop, the RG running of the neutrino mass matrix below the seesaw scale is described by the RG equation~\cite{rg}:
\begin{equation}
16\pi^2 \frac{\textrm{d}\kappa}{\textrm{d}t} = C (Y_l Y_l^\dagger)\kappa +C \kappa (Y_l Y_l^\dagger) + \xi \kappa \label{eq27}
\end{equation}
where $\kappa$ denotes the effective dimension five neutrino mass operator~\cite{weinberg}, $Y_l$ is the Yukawa coupling matrix for the charged leptons, $C = -\frac{3}{2}$ in the SM and $\xi$ denotes the contribution from gauge interactions. The renormalization scale $\mu$ enters through $t = \textrm{ln}(\mu / \mu_o)$. From Eq. (\ref{eq27}) one can see that in the flavor basis where the charged lepton Yukawa coupling matrix is diagonal, the radiative corrections to each element of the effective neutrino mass matrix are multiplicative, so that a zero entry remains zero.\\
The situation changes at energies larger than the mass scale of the lightest right-handed neutrino. Above this threshold, the neutrino Yukawa couplings $Y_\nu$ also contribute to the RG equations. Above the highest seesaw scale, the effective neutrino mass matrix is defined as 
\begin{equation}
M_\nu = -\frac{v^2}{2} Y_\nu M_R^{-1} Y_\nu^T
\end{equation}
where $v$ denotes the Higgs doublet vacuum expectation value (VEV). Between the mass thresholds, the singlet neutrinos are successively integrated out which leads to modifications in running \footnote{For details about the procedure of integrating out the right-handed neutrinos see Refs.~\cite{rgseesaw,schmidt2}}. In the full type-I+II seesaw scenario, the running of the effective neutrino mass matrix $M_\nu$ above and between the seesaw scales is given by the running of following three different contributions to $M_\nu$:
\begin{align}
M_\nu^{(1)} & = -\frac{v^2}{4} \kappa \\
M_\nu^{(2)} & = -\frac{v^2}{2} Y_\nu M_R^{-1} Y_\nu^T \\
M_\nu^{(3)} & = \frac{v^2}{2} \Lambda M_\triangle^{-2} Y_\triangle
\end{align} 
where $\Lambda, M_\triangle >> v$. The one loop $\beta$-functions for the effective neutrino mass matrix in various effective theories can be summarized as~\cite{schmidt2}
\begin{equation}
16\pi^2 \frac{\textrm{d}M_\nu^{(i)}}{\textrm{d}t} = [C_l Y_l Y_l^\dagger + C_\nu Y_\nu Y_\nu^\dagger + C_\triangle Y_\triangle Y_\triangle^\dagger]^T M_\nu^{(i)} + M_\nu^{(i)} [C_l Y_l Y_l^\dagger + C_\nu Y_\nu Y_\nu^\dagger + C_\triangle Y_\triangle Y_\triangle^\dagger] + \xi M_\nu^{(i)}
\end{equation}
where $M_\nu^{(i)}$ denotes any of the three contributions to the effective neutrino mass matrix.\\
For all the Yukawa coupling matrices realized in this work, the hermitian products $Y_k Y_k^\dagger$ ($k = l,\nu,\triangle$) which are relevant for the RG evolution of $M_\nu$ come out to be diagonal so that RG corrections are again multiplicative on the effective neutrino mass matrix elements leaving zero elements intact.
Thus, although, the values of neutrino masses and neutrino mixing parameters change due to RG corrections while running down from the seesaw scale to the electroweak scale, the correlations induced between neutrino masses and mixing parameters by texture zeros remain unchanged due to the stability of texture zeros.
\section{Summary}
The two texture zero neutrino mass matrices in the flavor basis have been fairly successful in describing neutrino masses and mixings. We have studied the implications of large effective Majorana neutrino mass for a class of two texture zero neutrino mass matrices. We found that Classes $B_1$, $B_2$, $B_3$ and $B_4$ all predict near maximal atmospheric neutrino mixing angle when supplemented with the assumption of large effective Majorana neutrino mass. Moreover, the near maximality of the atmospheric neutrino mixing angle is independent of the values of the solar and reactor neutrino mixing angles. Furthermore, we have shown how one can obtain such texture structures in the context of type-I+II seesaw using the discrete Abelian group $Z_3$. Assuming type-II seesaw dominance one can explain the hierarchy between the non-zero elements of the effective neutrino mass matrix. The hermitian products $Y_k Y_k^\dagger$ for various Yukawa coupling matrices realized in this work are all diagonal, leading to the stability of texture zeros against running from the seesaw to the electroweak scale. The assumption of large $|M_{ee}|$ is motivated by the fact that there are a number of forthcoming and presently ongoing experiments searching for neutrinoless double beta decay. These experiments are capable of confirming or ruling out large $|M_{ee}|$. 
\acknowledgements{
R. R. G. gratefully acknowledges the financial support provided by the University Grants Commission (UGC), Government of India, under the Dr. D. S. Kothari postdoctoral fellowship scheme.}


\begin{thebibliography}{99}

\bibitem{t2k} K. Abe et al. [T2K Collaboration], \textit{Phys. Rev. Lett.} \textbf{107}, 041801 (2011), arXiv:1106.2822 [hep-ex].

\bibitem{minos} P. Adamson et al. [MINOS Collaboration], \textit{Phys. Rev. Lett.} \textbf{107}, 181802 (2011), arXiv:1108.0015 [hep-ex].

\bibitem{dchooz} Y. Abe et al., [Double Chooz Collaboration], \textit{Phys. Rev. Lett.} \textbf{108}, 131801 (2012), arXiv:1112.6353 [hep-ex].

\bibitem{dayabay} F. P. An et al., [Daya Bay Collaboration], \textit{Phys. Rev. Lett.} \textbf{108}, 171803 (2012), arXiv:1203.1669 [hep-ex].

\bibitem{reno} J. K. Ahn et al., [RENO Collaboration], \textit{Phys. Rev. Lett.} \textbf{108}, 191802 (2012), arXiv:1204.0626 [hep-ex].

\bibitem{tbm} P. F. Harrison, D. H. Perkins and W. G. Scott, \textit{Phys. Lett.} \textbf{B 530}, 167 (2002), hep-ph/0202074; P. F. Harrison and W. G. Scott, \textit{Phys. Lett.} \textbf{B 535}, 163 (2002), hep-ph/0203209; Zhi-zhong Xing, \textit{Phys. Lett.} \textbf{B 533}, 85 (2002), hep-ph/0204049.

\bibitem{bm} F. Vissani, hep-ph/9708483; V. D. Barger, S. Pakvasa, T. J. Weiler and K. Whisnant, \textit{Phys. Lett.} \textbf{B 437}, 107 (1998), hep-ph/9806387; A. J. Baltz, A. S. Goldhaber and M. Goldhaber, \textit{Phys. Rev. Lett.} \textbf{81}, 5730 (1998), hep-ph/9806540.

\bibitem{gr1} A. Datta, F. S. Ling and P. Ramond, \textit{Nucl. Phys.} \textbf{B 671}, 383 (2003), hep-ph/0306002; Y. Kajiyama, M. Raidal and A. Strumia, \textit{Phys. Rev.} \textbf{D 76}, 117301 (2007), arXiv:0705.4559 [hep-ph]; L. L. Everett, A. J. Stuart, \textit{Phys. Rev.} \textbf{D 79}, 085005 (2009), arXiv:0812.1057 [hep-ph]; F. Feruglio, A. Paris, \textit{JHEP} \textbf{1103}, 101 (2011), arXiv:1101.0393 [hep-ph].

\bibitem{gr2} W. Rodejohann, \textit{Phys. Lett.} \textbf{B 671}, 267 (2009), arXiv:0810.5239 [hep-ph]; A. Adulpravitchai, A. Blum and W. Rodejohann, \textit{New J. Phys.} \textbf{11}, 063026 (2009), arXiv:0903.0531 [hep-ph].

\bibitem{hm}C. H. Albright, A. Dueck and W. Rodejohann, \textit{Eur. Phys. J.} \textbf{C 70}, 1099 (2010), arXiv:1004.2798 [hep-ph].

\bibitem{fgm} Paul H. Frampton, Sheldon L. Glashow and Danny Marfatia, \textit{Phys. Lett.} \textbf{B 536}, 79 (2002), hep-ph/0201008.
\bibitem{tz} Zhi-zhong Xing, \textit{Phys. Lett.} \textbf{B 530}, 159 (2002), hep-ph/0201151; Bipin R. Desai, D. P. Roy and Alexander R. Vaucher, \textit{Mod. Phys. Lett.} \textbf{A 18}, 1355 (2003), hep-ph/0209035; A. Merle, W. Rodejohann, \textit{Phys. Rev} \textbf{D 73}, 073012 (2006), hep-ph/0603111;  S. Dev, Sanjeev Kumar, S. Verma and S. Gupta, \textit{Nucl. Phys.} \textbf{B 784}, 103 (2007), hep-ph/0611313; G. Ahuja, S. Kumar, M. Randhawa, M. Gupta, S. Dev, \textit{Phys. Rev.} \textbf{D 76}, 013006 (2007), hep-ph/0703005; S. Kumar, \textit{Phys. Rev.} \textbf{D 84}, 077301 (2011), arXiv:1108.2137 [hep-ph]; D. Meloni, G. Blankenburg, \textit{Nucl. Phys.} \textbf{B 867}, 749 (2013), arXiv:1204.2706 [hep-ph]; W. Grimus, P. O. Ludl, \textit{J. Phys.} \textbf{G 40}, 055003 (2013) arXiv:1208.4515 [hep-ph]; Manmohan Gupta, Gulsheen Ahuja, \textit{Int. J. Mod. Phys.} \textbf{A, 27}, 1230033 (2012), arXiv:1302.4823 [hep-ph];  J. Liao, D. Marfatia, K. Whisnant, arXiv:1311.2639 [hep-ph]; D. Meloni, A. Meroni, E. Peinado, \emph{Phys. Rev.} {\bf D 89} (2014) 053009, arXiv:1401.3207 [hep-ph].

\bibitem{ourdegeneracies}S. Dev, S. Kumar, S. Verma and S. Gupta, \textit{Phys. Rev.} \textbf{D 76}, 013002 (2007), hep-ph/0612102.

\bibitem{xingtz} H. Fritzsch, Zhi-zhong Xing, S. Zhou, \textit{JHEP } \textbf{1109}, 083 (2011), arXiv:1108.4534 [hep-ph].

\bibitem{peinado} P. O. Ludl, S. Morisi, E. Peinado, \textit{Nucl. Phys.} \textbf{B 857}, 411 (2012), arXiv:1109.3393 [hep-ph].

\bibitem{zmlashin} E. I. Lashin and N. Chamoun, \textit{Phys. Rev.} \textbf{D 78}, 073002 (2008), arXiv:0708.2423 [hep-ph].

\bibitem{zerominor} L. Lavoura, \textit{Phys. Lett.} \textbf{B 609}, 317 (2005), hep-ph/0411232; E. I. Lashin, N. Chamoun, \textit{Phys. Rev.} \textbf{D 80}, 093004 (2009), arXiv:0909.2669 [hep-ph]; S. Dev, S. Verma, S. Gupta and R. R. Gautam, \textit{Phys. Rev.} \textbf{D 81}, 053010 (2010), arXiv:1003.1006 [hep-ph]; S. Dev, S. Gupta and R. R. Gautam, \textit{Mod. Phys. Lett.} \textbf{A 26}, 501 (2011), arXiv:1011.5587 [hep-ph];  T. Araki, J. Heeck and J. Kubo, \textit{JHEP} \textbf{1207}, 083 (2012), arXiv:1203.4951 [hep-ph]; Jiajun Liao, Danny Marfatia, Kerry Whisnant,  \textit{Phys. Rev.} \textbf{D 88}, 033011 (2013), arXiv:1306.4659 [hep-ph]; Weijian Wang, arXiv:1311.6944 [hep-ph]; Weijian Wang, arXiv:1402.6808 [hep-ph]. 

\bibitem{ourzm} S. Dev, Shivani Gupta, Radha Raman Gautam and Lal Singh, \textit{Phys. Lett.} \textbf{B 706}, 168 (2011), arXiv:1111.1300 [hep-ph].

\bibitem{hybrid} S. Kaneko, H. Sawanaka and M. Tanimoto, \textit{JHEP} \textbf{0508}, 073 (2005), hep-ph/0504074; S. Dev, S. Verma and S. Gupta, \textit{Phys. Lett.} \textbf{B 687}, 53-56 (2010), arXiv:0909.3182 [hep-ph]; S. Dev, S. Gupta and R. R. Gautam, \textit{Phys. Rev.} \textbf{D 82}, 073015 (2010) arXiv:1009.5501 [hep-ph]; Ji-Yuan Liu, Shun Zhou, \textit{Phys. Rev.} \textbf{D 87}, 093010 (2013), arXiv:1304.2334 [hep-ph];  Weijian Wang, \textit{Eur. Phys. J.} \textbf{C 73}, 2551 (2013), arXiv:1306.3556 [hep-ph]; S. Dev, R. R. Gautam and Lal Singh, \textit{Phys. Rev.} \textbf{D 88}, 033008 (2013), arXiv:1306.4281 [hep-ph].

\bibitem{tec} S. Dev, R. R. Gautam and Lal Singh, \textit{Phys. Rev.} \textbf{D 87}, 073011 (2013) arXiv:1303.3092 [hep-ph].

\bibitem{grimus} W. Grimus, P. O. Ludl, \textit{Phys. Lett.} \textbf{B 700}, 356-361 (2011), arXiv:1104.4340 [hep-ph].

\bibitem{fogli} G. L. Fogli, E. Lisi, A. Marrone, A. Palazzo, \textit{Prog. Part. Nucl. Phys.} \textbf{57}, 742 (2006), hep-ph/0506083. 

\bibitem{jarlskog} C. Jarlskog, \textit{Phys. Rev. Lett.} \textbf{55}, 1039 (1985).

\bibitem{ndbdecay} F. T. Avignone III, S. R. Elliott, J. Engel, \textit{Rev. Mod. Phys.} \textbf{80}, 481 (2008), arXiv:0708.1033 [nucl-ex]; J. J. Gomez-Cadenas, J. Martin-Albo, M. Mezzetto, F. Monrabal, M. Sorel,  \textit{Riv. Nuovo  Cim.} \textbf{35}, 29 (2012), arXiv:1109.5515  [hep-ex]; S. M. Bilenky, C. Giunti, \textit{Mod. Phys. Lett.} \textbf{A 27}, 1230015, arXiv:1203.5250 [hep-ph].

\bibitem{ndbrodejohann} W. Rodejohann, \textit{Int. J. Mod. Phys.} \textbf{E, 20}, 1833 (2011), arXiv:1106.1334 [hep-ph].

\bibitem{cuoricino} C. Arnaboldi et al., [CUORICINO Collaboration], \textit{Phys. Lett.} \textbf{B 584}, 260 (2004).

\bibitem{cuore} C. Arnaboldi et al., \textit{Nucl. Instrum. Methods Phys. Res., Sect.} \textbf{A 518}, 775 (2004), hep-ex/0212053.

\bibitem{gerda} I. Abt et al., [GERDA Collaboration], hep-ex/0404039.

\bibitem{majorana} R. Gaitskell et al., [Majorana Collaboration], nucl-ex/0311013.

\bibitem{supernemo} A. S. Barabash [NEMO Collaboration], \textit{Czech. J. Phys.}, \textbf{52}, 567 (2002), nucl-ex/0203001.

\bibitem{exo} M. Danilov et al., \textit{Phys. Lett.} \textbf{B 480}, 12 (2000), hep-ex/0002003.

\bibitem{genius} H. V. Klapdor- Kleingrothaus, et al., \textit{Eur. Phys. J.} \textbf{A 12}, 147 (2001), hep-ph/0103062.

\bibitem{planck} P. A. R. Ade et al. [Planck Collaboration], arXiv:1303.5076 [astro-ph].

\bibitem{valledata} D. V. Forero, M. Tortola and J. W. F. Valle , \textit{Phys. Rev.} \textbf{D 86}, 073012 (2012), arXiv:1205.4018 [hep-ph].

\bibitem{dualmodels} J. Liao, D. Marfatia, K. Whisnant, \textit{Phys. Rev.} \textbf{D 89}, 013009 (2014), arXiv:1308.1368 [hep-ph].
 
\bibitem{grimussym} W. Grimus, A. S. Joshipura, L. Lavoura and M. Tanimoto,  \textit{Eur. Phys. J.} \textbf{C 36}, 227 (2004), hep-ph/0405016.

\bibitem{seesaw1} P. Minkowski, \textit{Phys. Lett.} \textbf{B 67}, 421 (1977); T. Yanagida, \textit{Proceedings of the Workshop on the Unified Theory and the Baryon Number in the Universe} (O. Sawada and A. Sugamoto, eds.), KEK, Tsukuba, Japan, 1979, p. 95: M. Gell-Mann, P. Ramond, and R. Slansky, \textit{Complex spinors and unified theories in supergravity} (P. Van Nieuwenhuizen and D. Z. Freedman, eds.), North Holland, Amsterdam, 1979, p.315; R. N. Mohapatra and G. Senjanovic, \textit{Phys. Rev. Lett.} \textbf{44}, 912 (1980).

\bibitem{seesaw2} W. Konetschny and W. Kummer, \textit{Phys. Lett.} \textbf{B 70}, 433 (1977); T. P. Cheng and L. F. Li, \textit{Phys. Rev.} \textbf{D 22}, 2860 (1980); J. Schechter and J. W. F. Valle, \textit{Phys. Rev.} \textbf{D 22}, 2227 (1980); G. Lazarides Q. Shafi and C. Wetterich, \textit{Nucl. Phys.} \textbf{B 181}, 287 (1981); R. N. Mohapatra and G. Senjanovic, \textit{Phys. Rev.} \textbf{D 23}, 165 (1981).

\bibitem{grimustz} Walter Grimus, Luis Lavoura, \textit{J. Phys.} \textbf{G 31}, 693-702 (2005), hep-ph/0412283.

\bibitem{ourtzsym} S. Dev, Shivani Gupta, R. R. Gautam, \textit{Phys. Lett.} \textbf{B 701}, 605-608 (2011), arXiv:1106.3451 [hep-ph].

\bibitem{joshipuratz} M. Hirsch, A. S. Joshipura, S. Kaneko and J. W. F. Valle, \textit{Phys. Rev. Lett.} \textbf{99}, 151802 (2007), hep-ph/0703046.

\bibitem{schmidt} W. Rodejohann, M. A. Schmidt, \textit{Phys. Atom. Nucl.} \textbf{69}, 1833 - 1841 (2006), hep-ph/0507300.

\bibitem{grimusreview} W. Grimus, hep-ph/0612311.

\bibitem{matriplet} E. Ma, U. Sarkar, \emph{Phys. Rev. Lett.} {\bf 80} (1998) 5716, hep-ph/9802445.

\bibitem{rg} P. H. Chankowski, Z. Pluciennik, \textit{Phys. Lett.} \textbf{B 316}, 312 (1993), hep-ph/9306333; K. S. Babu, C. N. Leung, J. Pantaleone, \textit{Phys. Lett.} \textbf{B 319}, 191 (1993), hep-ph/9309223; S. Antusch, M. Drees, J. Kersten, M. Lindner, M. Ratz, \textit{Phys. Lett.} \textbf{B 519}, 238 (2001), hep-ph/0108005.

\bibitem{weinberg} S. Weinberg, \textit{Phys. Rev. Lett.} \textbf{43}, 1566 (1979).

\bibitem{rgseesaw} S. F. King, N. N. Singh, \textit{Nucl. Phys.} \textbf{B 591}, 3 - 25 (2000), hep-ph/0006229; S. Antusch, J. Kersten, M. Lindner, M. Ratz, \textit{Phys. Lett.} \textbf{B 538}, 87 (2002), hep-ph/0203233; S. Antusch, J. Kersten, M. Lindner, M. Ratz, M. A. Schmidt, \textit{JHEP} \textbf{0503}, 024 (2005), hep-ph/0501272; Shamayita Ray, \textit{Int. J. Mod. Phys.} \textbf{A 25}, 4339 - 4384 (2010), arXiv:1005.1938 [hep-ph]. 

\bibitem{schmidt2} M. A. Schmidt, \textit{Phys. Rev.} \textbf{D 76}, 073010 (2007), erratum-ibid. \textbf{D 85}, 099903 (2012), arXiv:0705.3841 [hep-ph].
\end{thebibliography}
\end{document}